\documentclass[
 reprint,
superscriptaddress,
 amsmath,amssymb,
 aps,
]{revtex4-2}

\usepackage{graphicx}
\usepackage{dcolumn}
\usepackage{bm}
\usepackage{algpseudocode}
\usepackage{algorithm}
\usepackage[mathlines]{lineno}
\usepackage{verbatim} 
\usepackage{float}
\usepackage{ulem}
\usepackage{hyperref}

\newcommand{\avg}[1]{\langle #1 \rangle}

\usepackage{xcolor}
\usepackage{ulem}

\begin{document}

\preprint{APS/123-QED}

\title{Generalized Lotka-Volterra systems with quenched random interactions and saturating nonlinear response}

\newcommand{\DFA}{Department of Physics and Astronomy ``Galileo Galilei'', University of Padova, Italy}
\newcommand{\INFN}{INFN, Padova Division, Italy}
\newcommand{\NBFC}{National Biodiversity Future Center, Palermo, Italy}
\newcommand{\PNC}{Padova Neuroscience Center, University of Padova, Italy}

\author{Marco Zenari}
\email{marco.zenari.2@phd.unipd.it}
\affiliation{\DFA}
\affiliation{\PNC}

\author{Francesco Ferraro}
\affiliation{\DFA}
\affiliation{\INFN}
\affiliation{\NBFC}

\author{Sandro Azaele}
\affiliation{\DFA}
\affiliation{\INFN}
\affiliation{\NBFC}

\author{Amos Maritan}%
\affiliation{\DFA}
\affiliation{\INFN}
\affiliation{\NBFC}

\author{Samir Suweis}
\affiliation{\DFA}
\affiliation{\PNC}
\affiliation{\INFN}

\begin{abstract}
The generalized Lotka-Volterra (GLV) equations with quenched random interactions have been extensively used to investigate the stability and dynamics of complex ecosystems. However, the standard linear interaction model suffers from pathological unbounded growth, especially under strong cooperation or heterogeneity. This work addresses that limitation by introducing a Monod-type saturating nonlinear response into the GLV framework. Using Dynamical Mean Field Theory, we derive analytical expressions for the species abundance distribution in the Unique Fixed Point phase and show the suppression of unbounded dynamics. Numerical simulations reveal a rich dynamical structure in the Multiple Attractor phase, including a transition between high-dimensional chaotic and low-volatility regimes, governed by interaction symmetry. These findings offer a more ecologically realistic foundation for disordered ecosystem models and highlight the role of nonlinearity and symmetry in shaping the diversity and resilience of large ecological communities.

\end{abstract}

\maketitle

\section{Introduction}

The study of the stability and the emergent properties of large ecological communities characterized by random interactions has evolved significantly since May's seminal contribution \cite{May1972}. A plethora of analytical tools borrowed from statistical physics have enabled significant progress in understanding the properties of fixed points and their stability in large ecosystems with disordered couplings. These techniques include random matrix theory \cite{Mehta2004, livan2018introduction},
replica and cavity methods \cite{Diederich1989,Biscari1995,Oliveira2000}, and dynamic generating functionals \cite{Dominicis1978,Martin1973,Sommers1988}, which were originally developed for describing the properties spin systems both in and out of equilibrium \cite{Mezard1993}.

In particular, several studies have focused on the stability of random assemblies of a large number of species by analyzing the eigenvalue spectra of random Jacobian matrices \cite{Allesina2012,Suweis2013,allesina2015stability,gravel2016stability,stone2018feasibility,barbier2018generic,pigani2022delay}, also known in ecology as the community matrix. However, this approach not only does not take into account the possibility of asymptotic states other than fixed points but usually also considers the Jacobian independent of the stationary species abundances $\vec{x}^*$ (but see \cite{Gibbs2018, poley2024eigenvalue}).
A complementary approach is to study species population dynamics using generalized Lotka-Volterra equations (GLV) with quenched random disorder (QGLV) \cite{kessler2015generalized, bunin2017ecological, galla2018dynamically, Biroli2018, barbier2018generic}. These models typically exhibit three distinct dynamical regimes: global convergence to a unique fixed point (UFP), convergence or fluctuations between multiple equilibria (MA), or unbounded population growth (UG). The introduction of demographic noise to these models has revealed additional phases, such as a Gardner phase \cite{altieri2021properties}, while considering delayed or modular interactions, oscillatory phases appear \cite{saeedian2022effect,martinez2025stabilization,ferraro2025synchronization}.
In the UFP phase, the species abundance (SAD) can be calculated analytically resulting in a truncated Gaussian distribution \cite{bunin2017ecological, galla2018dynamically, Biroli2018, azaele2024generalized,ser2018ubiquitous,grilli2020}. A heterogeneous non-Gaussian SAD can be obtained in the MA phase \cite{pearce2020stabilization,arnoulx2024many} or if annealed rather than quenched interactions are considered \cite{suweis2024generalized,ferraro2025exact,zanchetta2025emergence}. Also, recent works have shown that when non-Gaussian or sparse interactions are considered, the SAD shape is directly linked to the distribution of interactions \cite{azaele2024generalized,tonolo2025generalized}.

In some of these cases, however, a UG phase is present, in which one or more species populations diverge to infinity. This ecologically unrealistic outcome is due to a pathology of the GLV model, which assumes that the per-capita growth rate of a given species depends linearly on the remaining populations. This can lead to a positive feedback when interactions are strongly cooperative or very heterogeneous, which can ultimately results in unbounded growth. To remedy this pathology of GLV equations with random coefficients, a possibility is to consider non-positive, non-Gaussian distributions for these coefficients \cite{rossberg2013food,cockrell2024self,azaele2024generalized}. Alternatively, non-linear functional responses can be introduced. This, moreover, makes models more ecologically realistic and, in fact, several functional responses have been proposed in population dynamics models.

Functional responses, which describe how a consumer's feeding rate changes with resource density, play a fundamental role in understanding species interactions and ecosystem dynamics \cite{Holling1965,palamara2021stochastic}. Holling classified these responses into three main types: Type I (linear), Type II (saturating), and Type III (sigmoidal) \cite{Holling1965}. Type I responses assume a linear relationship between resource density and consumption rate up to a maximum, while Type II responses show a decelerating intake rate that reaches an asymptote at high resource densities, reflecting handling time limitations. Type III responses exhibit a sigmoid curve, indicating low feeding rates at low resource densities followed by an accelerating phase before reaching saturation. 

For this reason, a recent work has extended the study of QGLV to include a nonlinear functional response \cite{sidhom2020ecological}. In this study, non-linear feedback mechanisms similar to the Holling Type II functional response were introduced into the GLV model, with a saturating feedback function. However, the complexity of this functional response posed significant analytical challenges, making it difficult to derive closed-form solutions for the GLV phases, and mainly numerical simulations were employed to explore the dynamics of the system. 

The Monod equation \cite{Monod1949}, originally developed to describe microbial growth kinetics, represents a special case of Type II functional response and has become widely used in ecological modeling due to its mathematical simplicity and biological realism. This equation describes how growth rate varies with resource concentration, approaching a maximum rate asymptotically, similar to the Michaelis-Menten kinetics in enzyme dynamics. 

Following \cite{suweis2024generalized}, which explored the role of the saturating function in the GLV dynamics with time-dependent interactions, we aim to investigate the dynamics and stability of the QGLV model with a Monod-type nonlinear response. Our approach begins with the derivation of the Dynamical Mean Field Theory (DMFT) for the system, which allows us to study the UFP phase and derive the corresponding SAD in a closed form. We then focus on the loss of stability and the emergence of chaos in the MA phase as a function of the mean, variance and covariance of the random interactions. Although the UG phase vanishes as a result of the saturating nonlinear response, the MA phase, as also seen in previous work \cite{sidhom2020ecological}, presents two distinct types of dynamics. These are characterized by using the indicators proposed in  \cite{sidhom2020ecological} and two additional quantities: the dimensionality of the dynamics and the Maximum Lyapunov exponent. 
Thus, our work provides a comprehensive understanding of the different phases displayed by QGLV with saturating functional responses.

\section{Generalized Lotka-Volterra model with saturating response}
We consider a large ecological community composed of $N$ species. We denote their abundances or biomass densities with $x_i(t)$, with $i=1,\dots,N$. We assume that the dynamics of the abundances follow the GLV equations with a nonlinear interaction term $J(x)$
\begin{equation}
    \dot{x}_i = r_i x_i\left[1-x_i/k_i+\sum_{j \neq i} \alpha_{ij}J(x_j)\right] +\lambda_i.
    \label{eq:glv}
\end{equation}
The parameters $r_i$ are the per-capita growth rates of the species, while $k_i$ are their carrying capacities. For simplicity, we consider these parameters equal across species and rescale time and abundances, setting them to unity, i.e., $r_i=1$ and $k_i=1$ for all $i=1,...,N$. A small migration term $\lambda_i$ is also introduced, and from now on, it will also set equal for all species, that is, $\lambda_i=\lambda$ for $i=1,...,N$.
The addition of a migration term allows the system to reach a stationary state at long times for all values of the interaction parameters, by preventing a dynamical slowdown. This is often referred to as ``aging'' in the disordered-systems literature, although this terminology is somewhat non-standard here, as the slowdown originates from near-extinction events rather than from a rugged energetic landscape \cite{roy2019numerical,altieri2021properties,arnoulx2023aging}. 

The function $J(x)$ saturates at large values of its argument and is introduced to capture the nonlinear interactions characteristic of biological systems. Placing this function inside the summation, rather than following the usual formulation of functional responses, enhances analytical tractability at the expense of certain biological constraints, such as clone consistency~\cite{ansmann2021building}, that is, invariance under relabeling. It is straightforward to show that its introduction constrains the abundances of all species to remain bounded at all times. Thus, introducing a nonlinear saturating interaction avoids a UG phase.

For definiteness, we consider Monod-type response $J(x)$ \cite{Monod1949}, that is also similar to the Holling type II  ~\cite{Holling1965} functional response 
\begin{equation}
    J(x) = \frac{ax}{1+a h x},
\end{equation}
where $a$ is the consumption rate and $h$ is the handling time. In what follows, we set $a=1$ without loss of generality, as its effect can be absorbed into a redefinition of the interaction coefficients $\alpha_{ij}$ and the handling time $h$. We also assume a uniform handling time $h$ across all species. This response function behaves linearly in the small limit $x$ ($x\ll1$) and saturates to the value $1/h$ for large values of $x$. While the results that we obtain depend on our specific choice of the function $J(x)$, the general phenomenology depends only on the saturating character of the response.

The parameters $\alpha_{ij}$ specifying the interactions between the species composing a large community are prohibitively numerous and potentially unquantifiable. Following a standard approach in the study on large ecological communities, we take the interaction coefficients $\alpha_{ij}$ to be randomly distributed. Specifically, we take them from a distribution for which
\begin{align}
    \textrm{mean}(\alpha_{ij}) &= \mu/N \\
    \textrm{std}(\alpha_{ij}) &= \sigma/\sqrt{N} \\
    \textrm{corr}(\alpha_{ij},\alpha_{ji}) &= \gamma.
\end{align}
The specific choice of the distribution from which the parameters are drawn does not affect the resulting DMFT equations in the large $N$ limit, provided that higher-order cumulants are subleading so that only the first two cumulants contribute at leading order, as discussed in Refs.~\cite{Bunin2016, roy2019numerical, azaele2024generalized}. In simulations, parameters are drawn from a normal distribution.

\begin{figure*}
    \includegraphics[width=0.8\linewidth]{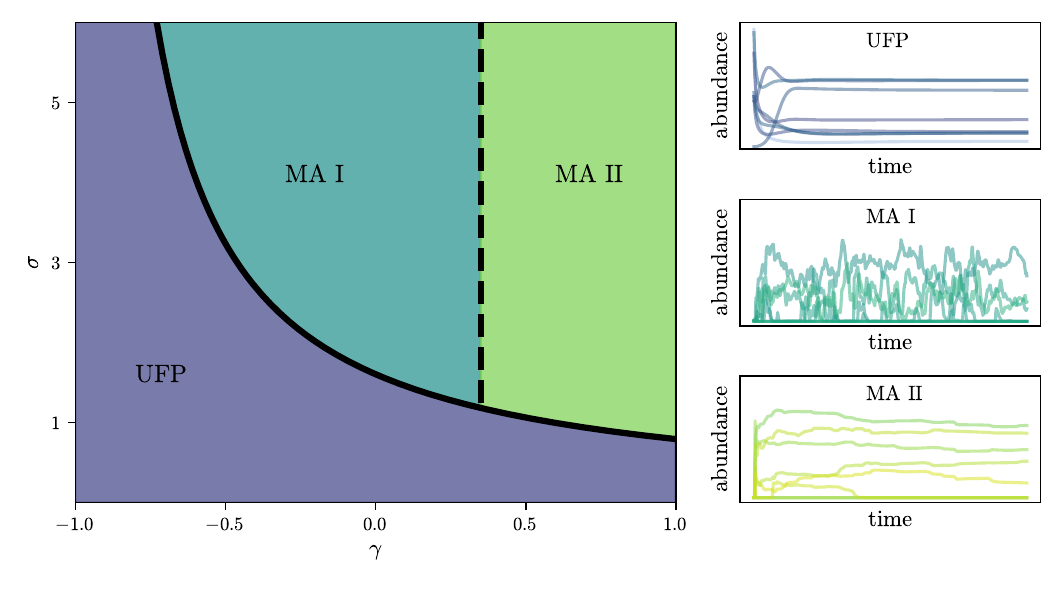}
    \caption{Qualitative behavior of the QGLV model with saturating nonlinear response, as a function of the strength of interactions $\sigma$ and the correlation parameter $\gamma$. The other parameters are set to $\mu = -3$, $h = 0.1$, $\lambda = 10^{-8}$. The insets show the trajectories of 8 random species among the $400$ used for the simulations for a total simulation time of $80$. The specific values of the parameters used are $\sigma = 5$ and $\gamma = -0.9, 0, 0.9$. The solid line marks the separation of between the Unique Fixed Point phase and the Multiple Attractors phase and is determined from the self-consistent condition Eq.(\ref{eq:critical_condition}). The dashed line marks the separation between the qualitatively different behaviors in the multiple attractors phase and is determined approximately with the order parameters shown in Fig. \ref{fig:panel4}.}
    \label{fig:panel1}
\end{figure*}

The qualitative behavior of the model is shown in Fig. \ref{fig:panel1}. Similarly to the standard GLV model, the system exhibits a UFP phase at low interaction strengths and undergoes a transition to the MA phase at a critical interaction strength that can be determined analytically. However, unlike the GLV equations with linear interactions, the inclusion of a saturating response function eliminates the nonphysical unbounded growth phase present in this model, which emerges for high cooperation and heterogeneity in the interactions, as can be easily demonstrated for a finite number of species $N$. The absence of this unbounded region enables a more comprehensive exploration of the MA phase, where, depending on the correlation parameter $\gamma$, two different transient chaotic behaviors are observed. We describe these two regions of the MA phase by quantifying their volatility numerically.

\section{Dynamical Mean-Field Theory}
In this section, we employ DMFT and map the multi-species, disordered system Eq. (\ref{eq:glv}) onto an effective single-species stochastic differential equation. To derive the DMFT equations, we follow a common approach used in the literature ~\cite{galla2018dynamically, galla2024generating}, which employs generating functionals. Starting from the GLV dynamic given by Eq. (\ref{eq:glv}), we define its generating functional as the Fourier transform of its measure
\begin{equation}
Z\left[{\Psi}\right]=\int_\text{paths} \mathcal{D}\left[\bold{x}\right] e^{i\sum_i\int dt x_i(t) \Psi_i(t)}, 
\end{equation}
where $\bold{\Psi}$ represent an external source field.

By performing the average of the generating functional over the disordered interaction parameters $\alpha_{ij}$ we show (see Appendix A) that this generating functional is equivalent, in the thermodynamic limit, to the one of the DMFT equation for an effective species $x(t)$ of the ecological community
\begin{multline}
\label{eq:DMFT}
    \dot{x}(t) = x(t)\bigg[1-x(t) + \gamma \sigma^2 \int_0^t dt' G(t,t') J(x(t')) \\+\mu Q(t) + \eta(t)\bigg].
\end{multline}
In this equation
\begin{equation}
    \label{eq:Q}
    Q(t) = \avg{J(x(t))},
\end{equation}
$\eta(t)$ is Gaussian noise, which has zero mean and correlations given by
\begin{equation}\label{eq:connector}
    \left\langle \eta(t) \eta(t') \right \rangle = \sigma^2 \left\langle J(x(t))J(x(t')) \right\rangle,
\end{equation}
and
\begin{equation}
    \label{eq:G}
    G(t, t') = \left\langle \frac{\delta J(x(t))}{\delta \eta (t')}\right \rangle.
\end{equation}
These averages are understood to be over realizations of Eq. (\ref{eq:DMFT}). Thus, the process defined by Eqs. (\ref{eq:DMFT}), (\ref{eq:Q}), (\ref{eq:connector}) and (\ref{eq:G}) is self-consistent.

We note that the DMFT equations could also be obtained by employing the so-called dynamical cavity method. This is discussed, for example, in the case of a linear response function in \cite{roy2019numerical}.

\section{Fixed point phase}
\subsection{Fixed point ansatz}
We begin by assuming that the dynamics defined by Eq. (\ref{eq:DMFT}) reach a fixed point in the long-time behavior and denote $x^* = \lim_{t \to \infty} x(t)$. With this assumption, the Gaussian noise $\eta(t)$ becomes a static Gaussian variable, as implied by Eq. (\ref{eq:connector}). Defining it as $\eta^* = \lim_{t \to \infty} \eta(t)$
\begin{equation}
    \eta^* = \sigma \sqrt{q} z,
\end{equation}
where
\begin{equation}\label{eq:q}
    q = \left \langle J(x(z))^2\right\rangle
\end{equation}
and $z$ is a zero-mean unit-variance Gaussian random variable. We additionally assume that the response function $G(t,t')$ at stationarity becomes time-translation invariant, $G(t, t')= G(t-t') = G(\tau)$, and define
\begin{equation}\label{eq:chi}
    \chi = \int d\tau G(\tau).
\end{equation}
We also define
\begin{equation}\label{eq:Q*}
    Q^* = \left \langle J(x(z)) \right \rangle.
\end{equation}

We find that the fixed points of Eq. (\ref{eq:DMFT}) are
\begin{equation}\label{eq:DMFT_stat}
x^*\left[1-x^*+\mu Q^* +\gamma \sigma^2 \chi J(x^*) +\eta^* \right] = 0.    
\end{equation}
The stable solution of Eq. (\ref{eq:DMFT_stat}) is given by
\begin{equation} \label{eq:stat_solution}
x^* = f(\xi) H[\xi],
\end{equation}
where $H[\xi]$ is the Heaviside function. The stability of the zero and non-zero fixed points is analyzed in the next section, see Eq. (\ref{eq:stability-zero-FP}), which justifies a posteriori the presence of the Heaviside function, see also \cite{bunin2017ecological,galla2018dynamically} for similar calculations. The argument $\xi$ is defined as
\begin{equation} \label{eq:xi}
\xi = 1 + \mu Q^* + \eta^*,
\end{equation}
and the function $f(\xi)$ is given by
\begin{equation} \label{eq:f_t}
f(\xi) = \frac{-2\xi}{h\xi - y - \sqrt{(h\xi - y)^2 + 4h\xi}},
\end{equation}
with $ y = 1 - \gamma \sigma^2 \chi$. Additional details on the derivation of this solution are provided in Appendix B. We note that in the limit $h\rightarrow 0$ Eq. (\ref{eq:stat_solution}) yields $f(\xi) = \xi/y$ which is the result obtained in \cite{galla2018dynamically}. The evaluation of $\chi$ can be carried out starting from Eq. (\ref{eq:chi}) and noting that, at stationarity, $\chi$ is the response of the system to a constant perturbation applied at all previous times~\cite{galla2024generating}. Thus, $\chi$ is obtained as the average over the realizations of $x^*=x(z)$ of the derivative of $J(x(z))$ with respect to $\eta^*= \sigma \sqrt{q} z$, 
\begin{equation}
    \chi = \int Dz \frac{1}{\sqrt{q}\sigma} \frac{\partial J(x(z))}{\partial z},
\end{equation}
where $Dz = \frac{1}{\sqrt{2\pi}}e^{-\frac{z^2}{2}} dz$ is the Gaussian measure. Computing the derivative, we obtain 
\begin{equation}
    \chi = \int _{\xi>0} d\xi \hspace{0.1cm} P_\xi(\xi) \frac{1}{(1+hf(\xi))^2}f'(\xi),
\end{equation}
with $f'(\xi)$ being the derivative respect to $\xi$ of Eq. (\ref{eq:f_t}).
\subsection{Species Abundance Distribution}
Starting from the stationary solution of DMFT given by Eq. (\ref{eq:DMFT_stat}) we derive the stationary probability distribution for $x^*$, which corresponds to the SAD. Eq. (\ref{eq:xi}) implies that $\xi$ is normally distributed as a Gaussian random variable $\mathcal{N}(1+\mu Q^*, \sigma^2 q)$. Thus, from Eq. \ref{eq:stat_solution}, we obtain
\begin{align}
    P_{\text{st}}(x^*) & = \langle \delta ( x^* -f(\xi)H(\xi))\rangle_{\xi}=\nonumber \\
    &=\phi\, \delta(x^*)+ P_{\text{surv}}(x^*)H(x^*)
    \label{stationary}
\end{align}
where the average is over $\xi$. The fraction of extinct species is given by
\begin{equation}
    \phi = \frac{1}{2}\text{erfc}\left({\frac{m}{\sqrt{2}\nu}}\right),
\end{equation}
while the distribution of non-extinct species distribution is:
\begin{equation} \label{eq:P_x}
 P_{\text{surv}}(x^*) = \frac{\exp{\left[-(f^{-1}(x^*)-m)^2/2\nu^2\right]}}{\sqrt{2\pi}\nu f'(f^{-1}(x^*))},
\end{equation}
where
\begin{equation}
    f^{-1}(x)= x - \gamma\sigma^2\chi J(x).
\end{equation}
The parameters $m = 1+\mu Q^*$ and $\nu = \sigma \sqrt{q}$ are to be computed self-consistently. Notice that $P_{\text{surv}}$ is normalized to $1-\phi$, the fraction of non-extinct species.

\begin{figure}
    \includegraphics[width=1\linewidth]{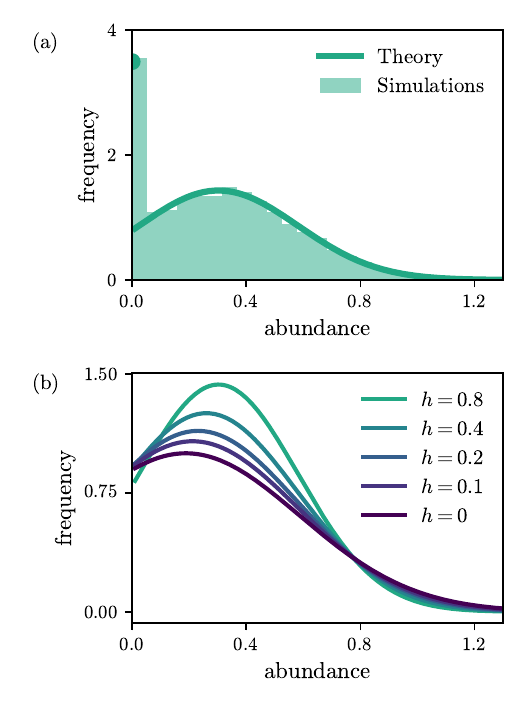}
\caption{
Species Abundance Distribution.(a) Comparison between the theoretical species abundance distribution and the histogram of the stationary samples obtained from $10$ simulations with $1000$ species. The values $x^*$ from the simulations are averaged over the last $5\%$ of the trajectories that last for a total simulation time of $100$. The used parameters are $\mu = -3$, $\sigma = 1$, $\gamma = 0$, $h=0.8$ and $\lambda = 10^{-8}$. The solid line represents the survival distribution $P_{\text{surv}}(x^*)$ while the point at $x^*=0$ is given by the sum of the probability of extinction $\phi$ and the fraction of non-extinct species with abundances contained within the first bin, both normalized by the bin length. (b) Comparison among the theoretical species abundance distributions for different values of $h$. Other parameters are fixed as $\mu = -3$, $\sigma = 1$, $\gamma = 0$ and $\lambda = 10^{-8}$.}
\label{fig:panel2}
\end{figure}

In the specific case $\gamma = 0$ we have $y=1$, $f^{-1}(x^*)=x^*$ and $f'(\xi) = 1$ and thus the resulting SAD is exactly a truncated Gaussian in Eq. (\ref{eq:P_x}), as obtained in the linear response case \cite{bunin2017ecological,galla2018dynamically}. For $\gamma\neq0$ the SAD is the pushforward of a Gaussian through $f^{-1}(x)$, but since $f^{-1}(x)$ is linear for large $x$, the SAD also decays in this case with a Gaussian tail. In Fig. \ref{fig:panel2}, we show the theoretical prediction together with the histogram of simulation samples for a representative choice of parameters. Results for other parameter choices are provided in Appendix \ref{app:Supp_Figures}.

\section{Loss of stability and emergence of chaos}

\subsection{Linear stability analysis}
In this section, we study the stability of the fixed point ansatz with a linear stability analysis following \cite{galla2018dynamically}. We add to the DMFT equation a small independent Gaussian white noise $\zeta(t)$
\begin{multline}
    \dot{x}(t) = x(t)[1-x(t) + \gamma \sigma^2 \int_0^t dt' G(t,t') J(x(t')) \\
    +\mu Q(t) + \eta(t) + \zeta(t)]
\end{multline}
and study the resulting deviations $\delta x(t) = x(t) - x^*$ from the stationary solution. At first order the noise is modified to $\eta(t) = \eta^* +\delta \eta(t)$, with
\begin{align} \label{eq:delta_eta}
    \langle \delta\eta(t) \delta\eta(t')\rangle =\sigma^2 \langle (J'^*)^2 \delta x(t) \delta x(t') \rangle \quad,
\end{align}
where $J'^* \equiv dJ(x)/dx\vert_{x=x^*}$.
The linearized DMFT equation is
\begin{multline} 
    \delta \dot{x} (t) = \delta x(t) [1-x^* +\gamma \sigma ^2 \chi J(x^*) +\mu Q^* +\eta^*] +\\
    + x^* [-\delta x(t) +\gamma \sigma^2 \int dt'[ \delta G(t,t') J(x(t')) +\\
    +G(t,t') \delta J(x(t')]+\mu \delta Q(t) + \delta \eta (t) + \zeta(t)], 
\end{multline}
and it is easy to show that $\delta G(t, t') = 0$ and $\delta Q(t) = 0$, starting from their definition and showing that it can be written as the average of an unperturbed observable multiplied by $\zeta$. Given the independence between the two and the fact that $\langle \zeta \rangle = 0 $, the terms mentioned above are zero.  

The dynamics of small perturbations around $x^* = 0$ thus satisfies
 \begin{equation}
     \delta \dot{x}(t) = \delta x(t) [1 + \mu Q^* + \eta^*]
     \label{eq:stability-zero-FP}
 \end{equation}
which shows that this fixed point is stable when $\xi = 1 + \mu Q^* + \eta^* < 0$, as anticipated.

The linear dynamics of perturbations around the non-zero fixed point satisfies instead
\begin{multline}
    \delta \dot{x}(t) = x^* [-\delta x(t) + \gamma \sigma ^2 \int dt' G(t, t') \delta J(x(t'))+
    \\+ \delta \eta (t) + \zeta(t)].
\end{multline}
We apply the Fourier transform on both sides, obtaining
\begin{multline}
i \omega \delta \Tilde{x}(\omega) = x^*[-\delta \Tilde{x}(\omega)+ \delta \Tilde{\eta}(\omega) + \Tilde{\zeta}(\omega)+\\
+\gamma \sigma ^2 \Tilde{G}(\omega) \delta \Tilde{x} (\omega) J'^* ],
\end{multline}
and solving for $\delta \Tilde {x}(\omega)$ we get
\begin{equation} \label{eq:perturbation}
    \delta \Tilde {x} (\omega) = A(\omega)^{-1}[\delta \Tilde{\eta}(\omega) + \Tilde{\zeta}(\omega) ],
\end{equation}
with 
\begin{equation}
    A(\omega) = \left[\frac{i\omega}{x^*}+1-\gamma \sigma^2 \Tilde{G}(\omega) J'^*\right].
\end{equation}
Since $\zeta$ is chosen to be a Gaussian white noise, we have
\begin{equation} \label{eq:delta_h}
\langle \Tilde{\zeta}(\omega) \Tilde{\zeta}(\omega') \rangle = \epsilon^2 \delta (\omega +\omega ') 2\pi,
\end{equation}
where $\epsilon$ is the strength of the Gaussian noise. Regarding $\langle \delta x(t) \delta x(t') \rangle$, since we are around the fixed point, it depends only on $t-t'$, and therefore
\begin{equation}\label{eq:delta_x}
    \langle \delta \Tilde{x}(\omega) \delta \Tilde{x}(\omega') \rangle = \epsilon^2 \delta (\omega +\omega ') 2\pi \delta \Tilde{C}(\omega),
\end{equation}
where $\delta \Tilde{C}(\omega)$ can be obtained from equations (\ref{eq:perturbation}), (\ref{eq:delta_h}), (\ref{eq:delta_eta}). Indeed, from eqs.(\ref{eq:delta_eta}) and (\ref{eq:perturbation}) we get
\begin{align}
    &\sigma^{-2}\langle \delta \Tilde{\eta}(\omega) \delta \Tilde{\eta}(\omega') \rangle = \langle (J'^*)^2\delta \Tilde{x}(\omega) \delta \Tilde{x}(\omega') \rangle =\nonumber\\
    &=D(\omega,\omega')\big[ \langle \delta \Tilde{\eta}(\omega) \delta \Tilde{\eta}(\omega') \rangle + \epsilon^2 \delta (\omega +\omega ') 2\pi\big],
    \label{consistency}
\end{align}
where
\begin{equation}
    D(\omega,\omega')\equiv \langle (A(\omega)A(\omega'))^{-1}(J'^*)^2\rangle.
\end{equation}
The last equation allows to determine the $\delta\tilde \eta$ correlation, which used in eq.(\ref{eq:perturbation}) leads to the final result:
\begin{equation}
    \delta \Tilde{C}(\omega) =  \frac{\langle \vert A(\omega)\vert^{-2}\rangle }{1-\sigma^2 D(\omega,-\omega)},
\end{equation}
 where the $\langle \cdot \rangle$ is an average over the distribution of $x^*$, with $x^*>0$. We are in particular interested in the long time response, and we therefore consider $\omega = \omega ' = 0$. In this case we obtain that
\begin{equation}\label{D}
    D(0,0) = \Big\langle \frac{(J'^*)^2}{(1-\gamma \sigma^2 \chi J'^*)^2}\Big\rangle,
\end{equation}
 where we have used that $\tilde G(0)=\chi$. $\delta \Tilde{C}(0)$ diverges if the following critical condition is satisfied
\begin{equation}\label{eq:critical_condition}
    1 = D(0,0) \sigma ^2,
\end{equation}
and the averages need to be computed with the stationary distribution of the surviving species 
\begin{equation}\label{normalization}
    P(x^*) = \frac{1}{1-\phi} P_{\text{surv}}(x^*),
\end{equation}
with $x^*>0$ and the pre-factor $1/(1-\phi)$ guarantees the normalization condition $\int_{x^*>0} dx^* P(x^*) =1$. The critical condition can be thus rewritten as
\begin{equation}\label{eq:stability_condition}
    1 = (1-\phi) \sigma^2 \int_{x^*>0} dx^* P(x^*) (J'^*)^2 \frac{1}{(1-\gamma \sigma^2 \chi J'^*)^2}.
\end{equation}
As expected, in the limit $h\rightarrow 0$ we obtain the same result of Ref. \cite{galla2018dynamically}.

The numerical validation of the critical-point prediction is discussed in the following section. Comparisons between theory and simulations for other choices of the model parameters are reported in Appendix~\ref{app:Supp_Figures}. 

As expected, in the limit of small handling times ($h \ll 1$), where $J(x) \sim x$, our results recover the phase diagram of the linear GLV model~\cite{bunin2017ecological}.

\begin{figure}[h!]
  \includegraphics[width=1\linewidth]{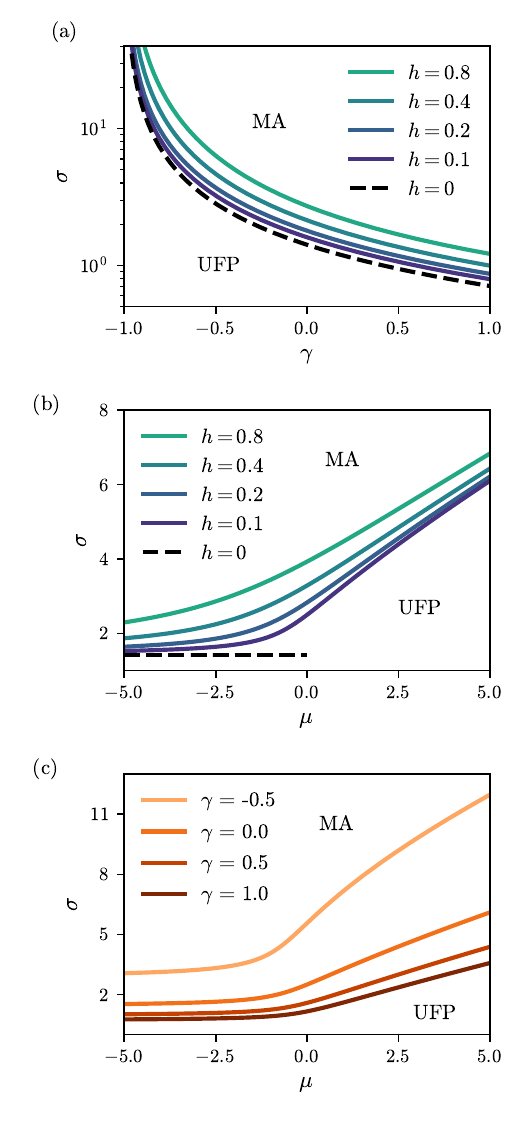}
  \caption{Different projections of the phase diagram. The lines separate the Unique Fixed Point phase (below) from the Multiple Attractor phase (above). (a) $\gamma$-$\sigma$ phase diagram with $\mu =-3$. Black dashed line is the result obtained in Ref. \cite{galla2018dynamically}. (b) $\mu$-$\sigma$ phase diagram with $\gamma =0$. Black dashed line is the result obtained in Ref. \cite{galla2018dynamically}. (c) $\mu$-$\sigma$ phase diagram with $h =0.1$.}
  \label{fig:panel3}
\end{figure}

The solution of Eq. (\ref{eq:stability_condition}) is shown in Fig. \ref{fig:panel3} as a function of the different control parameters of the model. As a function of $\gamma$, the qualitative behavior of the critical line between UFP and MA remains consistent with the standard GLV, but we observe that incorporating a saturating nonlinear response enlarges the region of stability of the system. Moreover, the greater the handling time, meaning that the nonlinear response saturates at lower abundances, the greater the stability. We also note that, differently from the case $J(x) = x$, the critical line delineating the UFP phase is not horizontal in the $\mu$-$\sigma$ space. The presence of a nonlinear saturating response allows us to explore in detail the region $\mu>0$ of cooperative communities, which is inaccessible in the linear GLV due to the nonphysical UG phase. As shown in panels b,c of Fig. \ref{fig:panel3}, our results indicate that the UFP region is larger in cooperative communities, which can thus be considered more stable than competitive ($\mu<0$) ones.

\subsection{Numerical characterization of simulations and validation}
\begin{figure*}[t]
    \includegraphics[width=0.8\linewidth]{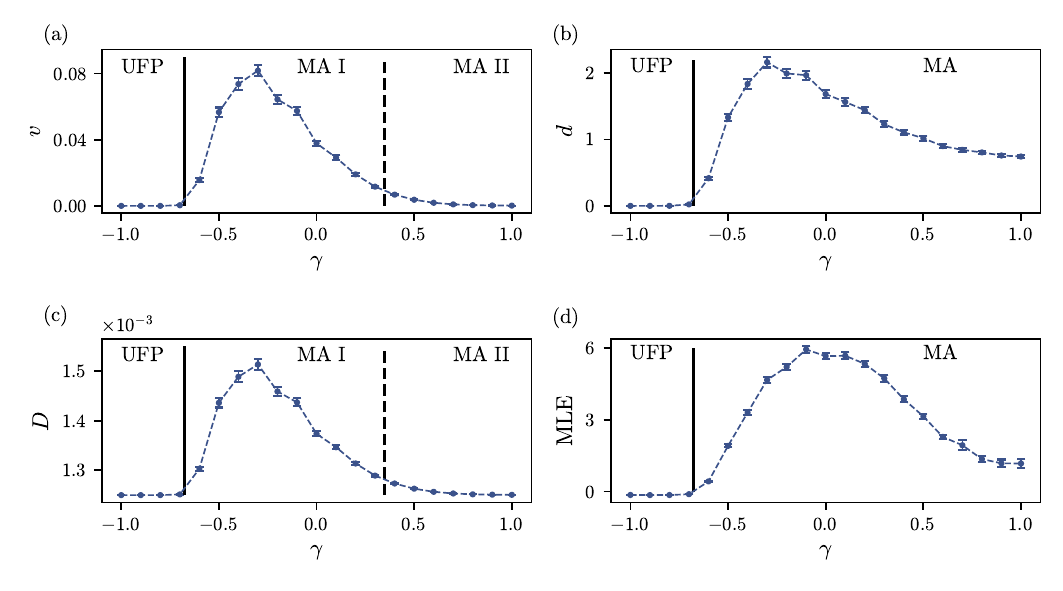}
    \caption{Comparison of the order parameter evaluated along a projection of the phase diagram, obtained fixing $\mu = -3$, $\sigma = 5$, $h = 0.1$ and varying $\gamma$. Each point is obtained as the average over $225$ realizations of the interaction matrix with $800$ species. The solid line marks the separation of between the Unique Fixed Point phase and the Multiple Attractors phase and is determined from the self-consistent condition Eq.(\ref{eq:stability_condition}). The dashed line marks the separation between the qualitatively different behaviors in the multiple attractors phase and is determined approximately.}
    \label{fig:panel4}
\end{figure*}
We performed extensive simulations to characterize the phases that are present beyond the unique fixed-point region. All code used to generate the results in this paper is publicly available at Ref.~\cite{repositoryQGLV}.
 We introduce two phenomenological order parameters to characterize the long-time behavior of the dynamics. Following Ref. \cite{sidhom2020ecological} we consider
\begin{equation} \label{eq:v_def}
    v = \frac{\langle\langle x_i(t)^2\rangle_T-\langle x_i(t)\rangle_T^2\rangle_N}{\langle \langle x_i(t)\rangle_T^2\rangle_N}
\end{equation}
and
\begin{equation}
    d = \frac{\langle\langle(x_i(t)-x_i'(t))^2\rangle_N\rangle_T}{\langle\langle x_i(t)\rangle_N^2\rangle_T}
\end{equation}
where $\langle \cdot \rangle_N$ is the average over the different populations, $\langle \cdot \rangle_T$ is the average over the last $5\%$ time steps of the trajectories and $x_i'$ is the population of the $i$-th species obtained by re-running the process with the same interaction matrix, but different initial conditions. The parameter $v$ characterizes the volatility of the dynamical trajectory at stationarity, while the parameter $d$ characterizes the distance between runs with the same interaction matrix and different initial conditions. In the UFP phase, we expect both $v = 0$ and $d = 0$ since the system invariably converges to a stable fixed point, which is unique and independent of the initial conditions. In the MA phase, we can have both $v>0$ and $d>0$ if the dynamics is volatile or $v= 0$ and $d>0$ if the system displays multiple and linearly stable fixed points. 

A similar characterization can be obtained by introducing more rigorous order parameters. The stability of the attractors can be taken into account by computing the Maximum Lyapunov Exponent 
\begin{equation}
    \text{MLE} = \lim_{t \to \infty} \lim_{d_0 \to 0} \frac{1}{t} \ln \frac{d_t}{d_0},
\end{equation}
where $d_0$ is a small perturbation of the trajectory at the initial time $t=0$ and $d_t $ is the separation of the unperturbed and perturbed trajectories after time $t$. We computed it numerically by implementing the algorithm proposed in ~\cite{benettin1980lyapunov}.

To quantify the volatility of the system, we introduce the dimension of activity, a measure commonly used in theoretical neuroscience to assess the number of degrees of freedom that are effectively engaged in the dynamics ~\cite{clark2023dimension}. It is defined as
\begin{equation}
    D = \frac{1}{N}\frac{\text{Tr}(C)^2}{\text{Tr}(C^2)},
\end{equation}
where $C$ is the covariance matrix of the species abundance time-series. The prefactor $1/N$ ensures that $D$ remains bounded between $0$ and $1$.
The numerical results for the introduced order parameters are shown in Fig. \ref{fig:panel4} as a function of the interaction symmetry $\gamma$. As expected, the parameters $v$ and $D$ exhibit a similar behavior, as well as the maximum Lyapunov exponent ($\text{MLE}$) and the parameter $d$. We note that, although a correspondence between $v$ and $D$ might be expected, the fact that they behave almost as linear transformations of each other in the MA~I phase is not obvious, as the former quantifies the average volatility of species abundances, whereas the latter captures the number and structure of principal modes in the dynamics. Exploring their relationship could provide a useful framework for investigating the dynamical behavior of the multiple-attractor phase in Generalized Lotka-Volterra models, which we leave for future work.
From now on, we will consider only the dimension of activity $D$ and the Maximum Lyapunov Exponent $\text{MLE}$ for our considerations. Starting from the fully asymmetric case $\gamma=-1$ and moving towards the fully symmetric case $\gamma = 1$, the system undergoes a sharp transition from the UFP phase to the MA phase, which is well predicted by Eq. (\ref{eq:stability_condition}). Within the MA phase, we identify a region where both $D$ and $\text{MLE}$ remain non-zero, characterizing the MA I phase (shown in Fig. \ref{fig:panel1}). As the system approaches a higher degree of symmetry, it displays less volatility, leading to a decrease in $D$, which eventually reaches zero, marking the transition to the MA II phase (also shown in Fig. \ref{fig:panel1}). We determine the transition between MA I and MA II numerically, identifying it as the value of $\gamma$ at which $D$ equals its value at the sharp transition from UFP to MA. We note that the exact values of \( D \) and the MLE are influenced by the choice of the migration term \( \lambda \). However, this dependence does not affect the validity of the arguments presented in this section. A numerical analysis of this effect is provided in Appendix~\ref{app:Supp_Figures}, Fig.~\ref{fig:supp_migration}.

\section{Discussion}
In this work, we have studied the effect of introducing a saturating nonlinear response in the generalized Lotka-Volterra model with quenched interactions so to remove the non-physical unbounded phase of the model.
In particular, we have chosen a Monod-like function, so to keep the model as simple as possible while ensuring that the system remains bounded. 
Interestingly, the simplicity of the model allows for analytical solutions of the SAD and for the fraction of extinct species that are close, but not the same, as the one obtained in the GLV model without saturating response. 
Similarly to the present work, in \cite{suweis2024generalized} the GLV equations were considered with the same type of saturating response. However, the interactions in \cite{suweis2024generalized} were modeled as annealed random variables with correlation time $\tau$. In that work, a single phase was found, while in the present work the parameter space separates into distinct phases with markedly different properties. Moreover, \cite{suweis2024generalized} did not consider a possible symmetry parameter $\gamma$ between reciprocal interactions as in the present work. A non-trivial effect of introducing the saturating response is the enrichment of the MA phase as a function of this interaction symmetry parameter $\gamma$, inducing the emergence of two types of dynamic behaviors: a region where chaotic dynamics with high species turnover is observed (MA I) and another region where species undergoes only small fluctuations in their population dynamics, with few (rare) switching events. The transitions between the two phases cannot be detected analytically through a linear stability analysis, but they have been characterized numerically through simulations by introducing two phenomenological order parameters: the volatility of the dynamic trajectory at stationarity $v$, and the distance between runs with the same
interaction matrix and different initial conditions $d$. The MA I phase has been defined as the one having  $v > 0$ and $d > 0$, while the MA II phase has $v = 0$ and $d > 0$. This result has also been validated using an alternative approach based on the Maximum Lyapunov Exponent (MLE),  
the classic measure of the sensitivity of a dynamical system to initial conditions and on the ``dimension of activity" $D$, a metric commonly used in theoretical neuroscience, which quantifies the number of degrees of freedom effectively involved in the dynamics. As expected, the parameters $v$ and $D$ exhibit a similar behavior, as well as MLE and $d$. 

The transition observed as the symmetry parameter $\gamma$ increases from $-1$ to $1$ offers significant insight into the role of interaction symmetry in shaping the system's dynamic regime. For large enough interactions heterogeneity ($\sigma$), our results reveal a sharp transition from the unique fixed-point (UFP) phase to the MA II with low volatility, passing through the MA I phase (high-dimensional and chaotic dynamic). This suggests that symmetry in the interaction matrix can serve as a control parameter, tuning the system between qualitatively distinct dynamical behaviors. This progressive reduction in dynamical complexity with increasing symmetry highlights a key organizing principle: higher symmetry leads to less volatile dynamics. Such findings resonate with broader themes in complex systems theory, where symmetry often governs the emergence of order and the suppression of chaos. The ability to delineate these phases and their transitions deepens our understanding of how structured interactions shape the repertoire of collective behaviors in high-dimensional systems.
Future research in this direction may employ the methodologies discussed in this work to analytically investigate the effect of the network topology into the interaction matrix, including trophic hierarchies, modular communities, or temporal variability, which at present have been explored mainly numerically \cite{barbier2018generic}. It would be valuable to extend the Dynamical Mean Field Theory framework to accommodate these complexities, possibly via perturbative expansions or machine learning-based  (e.g. Hamiltonian learning) surrogate models.

\section*{Acknowledgments}
F.F. acknowledges financial support under the National Recovery and Resilience Plan (NRRP), Mission 4, Component 2 Investment 1.4 - Call for tender No. 3138 of 16 December 2021, rectified by Decree n.3175 of 18 December 2021 of Italian Ministry of University and Research funded by the European Union - NextGenerationEU; Award Number: Project code CN00000033, Concession Decree No. 1034 of 17 June 2022 adopted by the Italian Ministry of University and Research, CUP C93C22002810006, Project title ``National Biodiversity Future Center - NBFC''.
S.S. acknowledges financial support under the National Recovery and Resilience Plan (NRRP), Mission 4, Component 2, Investment 1.1, Call for tender No. 104 published on 2.2.2022 by the Italian Ministry of University and Research (MUR), funded by the European Union – NextGenerationEU – Project Title: Anchialos: diversity, function, and resilience of Italian coastal aquifers upon global climatic changes – CUP C53D23003420001 Grant Assignment Decree n. 1015 adopted on 07/07/2023 by the Italian Ministry of Ministry of University and Research (MUR).

\bibliographystyle{unsrt}
\bibliography{biblio}

@article{barbier2018generic,
  title={Generic assembly patterns in complex ecological communities},
  author={Barbier, Matthieu and Arnoldi, Jean-Fran{\c{c}}ois and Bunin, Guy and Loreau, Michel},
  journal={Proceedings of the National Academy of Sciences},
  volume={115},
  number={9},
  pages={2156--2161},
  year={2018},
  publisher={National Academy of Sciences}
}

@article{May1972,
  title={Will a Large Complex System be Stable?},
  author={May, R. M.},
  journal={Nature},
  volume={238},
  pages={413},
  year={1972}
}

@article{Allesina2012,
  title={Stability Criteria for Complex Ecosystems},
  author={Allesina, S. and Tang, S.},
  journal={Nature},
  volume={483},
  pages={205},
  year={2012}
}

@book{Mehta2004,
  title={Random Matrices},
  author={Mehta, M. L.},
  series={Pure and Applied Mathematics},
  volume={142},
  year={2004},
  publisher={Elsevier Science Limited},
  address={Amsterdam}
}

@article{Gibbs2018,
  title={Effect of Population Abundances on the Stability of Large Random Ecosystems},
  author={Gibbs, T. and Grilli, J. and Rogers, T. and Allesina, S.},
  journal={Phys. Rev. E},
  volume={98},
  pages={022410},
  year={2018}
}

@article{Bunin2016,
  title={Interaction Patterns and Diversity in Assembled Ecological Communities},
  author={Bunin, G.},
  journal={arXiv:1607.04734},
  year={2016}
}

@article{Biroli2018,
  title={Marginally Stable Equilibria in Critical Ecosystems},
  author={Biroli, G. and Bunin, G. and Cammarota, C.},
  journal={New J. Phys.},
  volume={20},
  pages={083051},
  year={2018}
}

@article{Suweis2013,
  title={Emergence of Structural and Dynamical Properties of Ecological Mutualistic Networks},
  author={Suweis, S. and Simini, F. and Banavar, J. R. and Maritan, A.},
  journal={Nature},
  volume={500},
  pages={449},
  year={2013}
}

@article{Grilli2020,
  title={Macroecological Laws Describe Variation and Diversity in Microbial Communities},
  author={Grilli, J.},
  journal={Nat. Commun.},
  volume={11},
  pages={4743},
  year={2020}
}

@article{stone2018feasibility,
  title={The feasibility and stability of large complex biological networks: a random matrix approach},
  author={Stone, Lewi},
  journal={Scientific reports},
  volume={8},
  number={1},
  pages={1--12},
  year={2018},
  publisher={Nature Publishing Group}
}

@article{allesina2015stability,
  title={The stability--complexity relationship at age 40: a random matrix perspective},
  author={Allesina, Stefano and Tang, Si},
  journal={Population Ecology},
  volume={57},
  number={1},
  pages={63--75},
  year={2015},
  publisher={Wiley Online Library}
}

@article{pigani2022delay,
  title={Delay effects on the stability of large ecosystems},
  author={Pigani, Emanuele and Sgarbossa, Damiano and Suweis, Samir and Maritan, Amos and Azaele, Sandro},
  journal={Proceedings of the National Academy of Sciences},
  volume={119},
  number={45},
  pages={e2211449119},
  year={2022},
  publisher={National Academy of Sciences}
}

@book{Mezard1993,
  title={Spin Glass Theory and Beyond},
  author={M{\'e}zard, M. and Parisi, G. and Virasoro, M. A.},
  year={1993},
  publisher={World Scientific},
  address={Singapore}
}

@article{Diederich1989,
  title={Phase Transitions in Neural Networks with Random Asymmetric Couplings},
  author={Diederich, S. and Opper, M.},
  journal={Phys. Rev. A},
  volume={39},
  pages={4333},
  year={1989}
}

@article{Biscari1995,
  title={Statistical Mechanics of Random Replicator Networks},
  author={Biscari, P. and Parisi, G.},
  journal={J. Phys. A: Math. Gen.},
  volume={28},
  pages={4697},
  year={1995}
}

@article{palamara2021stochastic,
  title={The stochastic nature of functional responses},
  author={Palamara, Gian Marco and Capit{\'a}n, Jos{\'e} A and Alonso, David},
  journal={Entropy},
  volume={23},
  number={5},
  pages={575},
  year={2021},
  publisher={MDPI}
}

@article{Oliveira2000,
  title={Evolution in Species Space: Competition and Niche Selection},
  author={de Oliveira, V. M. and Fontanari, J.},
  journal={Phys. Rev. Lett.},
  volume={85},
  pages={4984},
  year={2000}
}

@article{Dominicis1978,
  title={Dynamics as a Substitute for Replicas in Systems with Quenched Random Impurities},
  author={De Dominicis, C.},
  journal={Phys. Rev. B},
  volume={18},
  pages={4913},
  year={1978}
}

@article{Martin1973,
  title={Statistical Dynamics of Classical Systems with Applications to Turbulence},
  author={Martin, P. C. and Siggia, E. D. and Rose, H. A.},
  journal={Phys. Rev. A},
  volume={8},
  pages={423},
  year={1973}
}

@article{Sommers1988,
  title={Theory of the Time Evolution of Neural Activity Patterns},
  author={Sommers, H. J. and Crisanti, A. and Sompolinsky, H. and Stein, Y.},
  journal={Phys. Rev. Lett.},
  volume={60},
  pages={1895},
  year={1988}
}

@article{Holling1965,
  title={The Functional Response of Predators to Prey Density and its Role in Mimicry and Population Regulation},
  author={Holling, C. S.},
  journal={Memoirs Entomol. Soc. Canada},
  volume={97},
  pages={5},
  year={1965}
}

@article{Monod1949,
  title={The Growth of Bacterial Cultures},
  author={Monod, J.},
  journal={Annu. Rev. Microbiol.},
  volume={3},
  pages={371},
  year={1949}
}

@article{galla2018dynamically,
  title={Dynamically evolved community size and stability of random Lotka-Volterra ecosystems (a)},
  author={Galla, Tobias},
  journal={Europhysics Letters},
  volume={123},
  number={4},
  pages={48004},
  year={2018},
  publisher={IOP Publishing}
}

@article{roy2019numerical,
  title={Numerical implementation of dynamical mean field theory for disordered systems: Application to the Lotka--Volterra model of ecosystems},
  author={Roy, Felix and Biroli, Giulio and Bunin, Guy and Cammarota, Chiara},
  journal={Journal of Physics A: Mathematical and Theoretical},
  volume={52},
  number={48},
  pages={484001},
  year={2019},
  publisher={IOP Publishing}
}

@article{bunin2017ecological,
  title={Ecological communities with Lotka-Volterra dynamics},
  author={Bunin, Guy},
  journal={Physical Review E},
  volume={95},
  number={4},
  pages={042414},
  year={2017},
  publisher={APS}
}

@article{arnoulx2024many,
  title={Many-species ecological fluctuations as a jump process from the brink of extinction},
  author={Arnoulx de Pirey, Thibaut and Bunin, Guy},
  journal={Physical Review X},
  volume={14},
  number={1},
  pages={011037},
  year={2024},
  publisher={APS}
}

@article{pearce2020stabilization,
  title={Stabilization of extensive fine-scale diversity by ecologically driven spatiotemporal chaos},
  author={Pearce, Michael T and Agarwala, Atish and Fisher, Daniel S},
  journal={Proceedings of the National Academy of Sciences},
  volume={117},
  number={25},
  pages={14572--14583},
  year={2020},
  publisher={National Academy of Sciences}
}

@article{martinez2025stabilization,
  title = {Stabilization of macroscopic dynamics by fine-grained disorder in many-species ecosystems},
  author = {Giral Mart\'{\i}nez, Juan and De Monte, Silvia and Barbier, Matthieu},
  journal = {Phys. Rev. E},
  volume = {112},
  issue = {3},
  pages = {034305},
  numpages = {8},
  year = {2025},
  month = {Sep},
  publisher = {American Physical Society},
}

@article{saeedian2022effect,
  title={Effect of delay on the emergent stability patterns in generalized Lotka--Volterra ecological dynamics},
  author={Saeedian, Meghdad and Pigani, Emanuele and Maritan, Amos and Suweis, Samir and Azaele, Sandro},
  journal={Philosophical Transactions of the Royal Society A},
  volume={380},
  number={2227},
  pages={20210245},
  year={2022},
  publisher={The Royal Society}
}

@article{gravel2016stability,
  title={Stability and complexity in model meta-ecosystems},
  author={Gravel, Dominique and Massol, Fran{\c{c}}ois and Leibold, Mathew A},
  journal={Nature communications},
  volume={7},
  number={1},
  pages={12457},
  year={2016},
  publisher={Nature Publishing Group UK London}
}

@article{sidhom2020ecological,
  title={Ecological communities from random generalized Lotka-Volterra dynamics with nonlinear feedback},
  author={Sidhom, Laura and Galla, Tobias},
  journal={Physical Review E},
  volume={101},
  number={3},
  pages={032101},
  year={2020},
  publisher={APS}
}

@article{altieri2021properties,
  title={Properties of equilibria and glassy phases of the random Lotka-Volterra model with demographic noise},
  author={Altieri, Ada and Roy, Felix and Cammarota, Chiara and Biroli, Giulio},
  journal={Physical Review Letters},
  volume={126},
  number={25},
  pages={258301},
  year={2021},
  publisher={APS}
}

@article{galla2024generating,
  title={Generating-functional analysis of random Lotka-Volterra systems: A step-by-step guide},
  author={Galla, Tobias},
  journal={arXiv preprint arXiv:2405.14289},
  year={2024}
}

@article{benettin1980lyapunov,
  title={Lyapunov characteristic exponents for smooth dynamical systems and for Hamiltonian systems; a method for computing all of them. Part 1: Theory},
  author={Benettin, Giancarlo and Galgani, Luigi and Giorgilli, Antonio and Strelcyn, Jean-Marie},
  journal={Meccanica},
  volume={15},
  pages={9--20},
  year={1980},
  publisher={Springer}
}

@article{clark2023dimension,
  title={Dimension of activity in random neural networks},
  author={Clark, David G and Abbott, LF and Litwin-Kumar, Ashok},
  journal={Physical Review Letters},
  volume={131},
  number={11},
  pages={118401},
  year={2023},
  publisher={APS}
}

@article{poley2024eigenvalue,
  title={Eigenvalue spectra of finely structured random matrices},
  author={Poley, Lyle and Galla, Tobias and Baron, Joseph W},
  journal={Physical Review E},
  volume={109},
  number={6},
  pages={064301},
  year={2024},
  publisher={APS}
}

@article{azaele2024generalized,
  title={Generalized dynamical mean field theory for non-Gaussian interactions},
  author={Azaele, Sandro and Maritan, Amos},
  journal={Physical Review Letters},
  volume={133},
  number={12},
  pages={127401},
  year={2024},
  publisher={APS}
}

@article{suweis2024generalized,
  title={Generalized lotka-volterra systems with time correlated stochastic interactions},
  author={Suweis, Samir and Ferraro, Francesco and Grilletta, Christian and Azaele, Sandro and Maritan, Amos},
  journal={Physical Review Letters},
  volume={133},
  number={16},
  pages={167101},
  year={2024},
  publisher={APS}
}

@article{ferraro2025exact,
  title={Exact solution of dynamical mean-field theory for a linear system with annealed disorder},
  author={Ferraro, Francesco and Grilletta, Christian and Maritan, Amos and Suweis, Samir and Azaele, Sandro},
  journal={Journal of Statistical Mechanics: Theory and Experiment},
  volume={2025},
  number={2},
  pages={023301},
  year={2025},
  publisher={IOP Publishing}
}

@article{ferraro2025synchronization,
  title={Synchronization and chaos in complex ecological communities with delayed interactions},
  author={Ferraro, Francesco and Grilletta, Christian and Pigani, Emanuele and Suweis, Samir and Azaele, Sandro and Maritan, Amos},
  journal={arXiv preprint arXiv:2503.21551},
  year={2025}
}

@article{zanchetta2025emergence,
  title={Emergence of ecological structure and species rarity from fluctuating metabolic strategies},
  author={Zanchetta, Davide and Gupta, Deepak and Moschin, Sofia and Suweis, Samir and Maritan, Amos and Azaele, Sandro},
  journal={PRX Life},
  volume={3},
  number={3},
  pages={033016},
  year={2025},
  publisher={APS}
}

@article{tonolo2025generalized,
  title={Generalized Lotka-Volterra model with sparse interactions: non-Gaussian effects and topological multiple-equilibria phase},
  author={Tonolo, Tommaso and Angelini, Maria Chiara and Azaele, Sandro and Maritan, Amos and Gradenigo, Giacomo},
  journal={arXiv preprint arXiv:2503.20887},
  year={2025}
}

@article{kessler2015generalized,
  title={Generalized model of island biodiversity},
  author={Kessler, David A and Shnerb, Nadav M},
  journal={Physical Review E},
  volume={91},
  number={4},
  pages={042705},
  year={2015},
  publisher={APS}
}

@article{ser2018ubiquitous,
  title={Ubiquitous abundance distribution of non-dominant plankton across the global ocean},
  author={Ser-Giacomi, Enrico and Zinger, Lucie and Malviya, Shruti and De Vargas, Colomban and Karsenti, Eric and Bowler, Chris and De Monte, Silvia},
  journal={Nature ecology \& evolution},
  volume={2},
  number={8},
  pages={1243--1249},
  year={2018},
  publisher={Nature Publishing Group UK London}
}

@article{livan2018introduction,
  title={Introduction to random matrices theory and practice},
  author={Livan, Giacomo and Novaes, Marcel and Vivo, Pierpaolo},
  journal={Monograph Award},
  volume={63},
  number={54},
  pages={914},
  year={2018}
}

@book{rossberg2013food,
  title={Food webs and biodiversity: foundations, models, data},
  author={Rossberg, Axel G},
  year={2013},
  publisher={John Wiley \& Sons}
}

@article{cockrell2024self,
  title={Self-organization of ecosystems to exclude half of all potential invaders},
  author={Cockrell, Cillian and O'Sullivan, Jacob D and Terry, J Christopher D and Nwankwo, Emmanuel C and Trachenko, Kostya and Rossberg, Axel G},
  journal={Physical Review Research},
  volume={6},
  number={1},
  pages={013093},
  year={2024},
  publisher={APS}
}

@article{arnoulx2023aging,
  title={Aging by near-extinctions in many-variable interacting populations},
  author={Arnoulx de Pirey, Thibaut and Bunin, Guy},
  journal={Physical Review Letters},
  volume={130},
  number={9},
  pages={098401},
  year={2023},
  publisher={APS}
}

@article{ansmann2021building,
  title={Building clone-consistent ecosystem models},
  author={Ansmann, Gerrit and Bollenbach, Tobias},
  journal={PLOS Computational Biology},
  volume={17},
  number={2},
  pages={e1008635},
  year={2021},
  publisher={Public Library of Science San Francisco, CA USA}
}

@misc{repositoryQGLV,
  author       = {Marco Zenari},
  title        = {QGLV-saturating-nonlinear-response},
  howpublished = {\url{https://github.com/MarcoZenari/QGLV-saturating-nonlinear-response}},
  note         = {GitHub repository},
  year         = {2025}
}

\clearpage
\onecolumngrid
\appendix
\section{Dynamical Mean Field Theory}
In this section we carry out the DMFT, or Functional analysis, of model (\ref{eq:glv}). We present here only the main points of the derivation, as it is standard in the literature of disordered systems. We refer the interested reader to \cite{galla2024generating} for a didactic presentation of the method for a similar setting to the one considered here, which presents all the calculations in full detail. The dynamical moment generating functional is defined as 
\begin{equation}
Z\left[\bold{\Psi}\right]=\int_{paths} \mathcal{D}\left[\bold{x}\right] e^{i\sum_i\int dt x_i(t) \Psi_i(t)},    
\end{equation}
where $\bold{\Psi}=(\Psi_1,...,\Psi_N)$ is an external source field and the integral is computed on the measure generated by the GLV dynamics. We can constrain the integral over the possible trajectories using a Dirac delta functional in its Fourier transform representation and obtain

\begin{equation}
Z\left[\bold{\Psi}\right]=\int \mathcal{D}[\bold{x}, \bold{\hat{x}}] \exp{\left(i\sum_i\int dt x_i(t) \Psi_i(t)\right )}\times \exp{\left(i\sum_i \int dt \hat{x}_i(t)\left(\frac{\dot{x}_i(t)}{x_i(t)} - \left[1-x_i(t)+\sum_{j \neq i} \alpha_{ij}J(x_j(t)) + \zeta(t)\right]\right)\right)} .
\end{equation}

 Included in the measure $\mathcal{D}[\bold{x}, \bold{\hat{x}}]$ is a Jacobian that guarantees that $Z\left[\bold{\Psi}=0\right]=1$. Its presence will be canceled at the end of the calculation, when we derive the actual stochastic single-species equation. 
Note that we have set $\lambda = 0$ and introduced a small field $\zeta(t)$ used to compute the response functions that will be set to $0$ at the end of the computation. Now we can perform the average over the quenched disorder due to the parameters $\alpha_{ij}$, that are distributed as
\begin{equation}
\alpha_{ij} = \frac{\mu}{N} + \frac{\sigma}{\sqrt{N}}z_{ij} \hspace{1cm}
\alpha_{ji} = \frac{\mu}{N} + \frac{\sigma}{\sqrt{N}}z_{ji},
\end{equation}
where $z_{ij}$ are normally distributed random variables for which $\overline{z_{ij}}=0$, $\overline{z_{ij}^2} = 1$ and $\overline{z_{ij}z_{ji}}=\gamma$.
The averaged over disorder functional becomes
\begin{equation}\label{eq:Z_averaged}
    \overline{Z[\bold{\Psi}]} = \int \mathcal{D}[P, Q, L, C, K, \hat{P}, \hat{Q}, \hat{L}, \hat{C}, \hat{K}] \exp{\left(N(\Psi + \Phi +\Omega)\right)},
\end{equation}
where we have introduced the order parameters
\begin{align}
    P(t) &= i \frac{1}{N}\sum_i \hat{x}_i(t)\\
    Q(t) &= \frac{1}{N} \sum_j J(x_j(t))\\
    C(t,t') &= \frac{1}{N} \sum_j J(x_j(t))J(x_j(t'))\\
    L(t,t') &= \frac{1}{N} \sum_i \hat{x}_i(t)\hat{x}_i(t')\\
    K(t,t') &= \frac{1}{N} \sum_i \hat{x}_i(t') J(x_i(t))
\end{align}
using the Dirac delta representation, e.g. for $L(t,t')$
\begin{equation}
    1 = \int \mathcal{D}[L, \hat{L}] \exp{\left(iN\int dt\hspace{0.1cm} dt' \hat{L}(t, t')\left(L(t,t')- \frac{1}{N} \sum_i \hat{x}_i(t)\hat{x}_i(t')\right)\right)}.
\end{equation}
In the argument of the exponential of Eq. (\ref{eq:Z_averaged}) we have pointed out the dependence on $N$ and we have that $\Psi$ comes from the introduction of the order parameters, $\Omega$ depends on the details of the microscopic time evolution and $\Phi$ comes from the average over the disorder:
\begin{equation}
    \Psi = i \int dt \int dt' \left[\hat{L}(t,t')L(t,t') + \hat{K}(t,t') K(t,t')+ \hat{C}(t,t')C(t,t') \right] + i\int dt \left[\hat{P}(t)P(t)+\hat{Q}(t) Q(t)\right]
\end{equation}
\begin{multline}
\Omega = \frac{1}{N}\sum_i \log \Biggl[
\int \mathcal{D}[x_i, \hat{x}_i] p_0^i(x_i(0))
  \exp\biggl(i\int dt\, J(x_i(t)) \Psi_i(t)\biggr)
  \exp\biggl(i\int dt\, \hat{x}_i(t)
      \bigl[\tfrac{\dot{x}_i(t)}{x_i(t)}-[1-x_i(t)+\zeta(t)]\bigr]\biggr)
\\
  \times \exp\biggl(-i\int dt\,dt'\,
     \bigl[\hat{L}(t,t') \hat{x}_i(t)\hat{x}_i(t')
     + \hat{C}(t,t') J(x_i(t))J(x_i(t'))
     + \hat{K}(t,t') \hat{x}_i(t') J(x_i(t))\bigr]\biggr)
\\
  \times \exp\biggl(-i\int dt\, \bigl[\hat{Q}(t) J(x_i(t))\bigr]\biggr)
  \times \exp\biggl(-i\int dt\, \bigl[\hat{P}(t) i\hat{x}_i(t)\bigr]\biggr)
\Biggr]
\end{multline}

\begin{equation}
    \Phi = -\frac{1}{2} \sigma^2 \int dt \hspace{0.1cm}dt' \left[L(t,t') C(t,t') +\gamma K(t,t') K(t',t) \right] - \mu \int dt P(t) Q(t),
\end{equation}
where $p_0^i(0)$is the distribution from which the initial values $x_i$ are sampled. Now we are ready to take the $N\rightarrow \infty$ limit and compute $\overline{Z[\bold{\Psi}]}$ with saddle point. From the optimization of the argument of the exponential of Eq. (\ref{eq:Z_averaged}) we obtain the following relations:
\begin{align}   
    \mu Q(t) &= i \hat{P}(t)\\
    i \hat{Q}(t) &= \mu P(t) \\
    \frac{1}{2}\sigma^2 C(t,t') &= i \hat{L}(t,t')\\
    \frac{1}{2} \sigma ^2 L(t,t') &= i \hat{C}(t,t')\\
    i\hat{K}(t,t') &= \frac{1}{2}\gamma \sigma^2 K(t,t')\\
    P(t) &= \lim_{N\to \infty}\frac{1}{N}\sum_i \langle i \hat{x}_i(t) \rangle_\Omega \\
    Q(t) &= \lim_{N\to \infty}\frac{1}{N}\sum_i \langle J(x_i(t)) \rangle_\Omega \\
    L(t, t') &= \lim_{N\to \infty}\frac{1}{N}\sum_i \langle  \hat{x}_i(t) \hat{x}_i(t') \rangle_\Omega \\
    C(t, t') &= \lim_{N\to \infty}\frac{1}{N}\sum_i \langle J(x_i(t))J(x_i(t')) \rangle_\Omega \\
    K(t, t') &= \lim_{N\to \infty}\frac{1}{N}\sum_i \langle i \hat{x}_i(t') J(x_i(t)) \rangle_\Omega.
\end{align}  
From the normalization constraint of the functional $Z$ it is possible to show that $P(t)=0$ $\forall t$ and $L(t,t')=0$ $\forall t,t'$.  Finally, introducing $G(t,t')=-iK(t,t')$ the effective dynamical generating functional is
\begin{multline}
    Z_{eff} = \int D[x, \hat{x}] P_0(x(0)) \times \exp {\left(i \int dt \hat{x}(t) \left(\frac{\dot{x}(t)}{x(t)}-[1-x(t)+ \zeta(t)+\mu Q(t)]\right)\right)} \times \\
    \times \exp{\left(-\sigma^2 \int dt \hspace{0.1cm} dt' \left[\frac{1}{2}C(t,t') \hat{x}(t) \hat{x}(t') +i\gamma G(t,t') J(x(t)) \hat{x}(t') \right]\right)},
\end{multline}
which is the generating functional of the effective dynamics
\begin{equation}
    \dot{x}(t) = x(t)[1-x(t) +\gamma \sigma ^2 \int dt' G(t,t') J(x(t')) + \mu Q(t) +\eta(t) +\zeta(t)],
\end{equation}
with 
\begin{align}
    C_\eta(t,t') &= \langle \eta(t) \eta(t') \rangle_x =\sigma^2 \left\langle J(x(t))J(x(t')) \right\rangle_x, \\
    G(t, t') &= \left\langle \frac{\delta J(x(t))}{\delta \zeta(t')} \right\rangle_x, \\
    Q(t) &= \left\langle J(x(t)) \right\rangle_x. 
\end{align}.

\section{Stable solution of stationary equation}
Starting from Eq. (\ref{eq:DMFT_stat}) we note that there is a trivial solution $x^*=0$, and possible other solutions can be obtained solving the equation
\begin{equation}
    1-x^* +\gamma \sigma^2 \chi J(x^*) +\mu Q^* + \eta ^* = 0,
\end{equation}
with $J(x^*) = x^*/(1+hx^*)$. Finally we have a second degree polynomial equation in $x^*$. The solutions are 
\begin{equation}\label{eq:second_order_sol}
    x^* = \frac{h\xi-y \pm \sqrt{(h\xi-y)^2+4h\xi}}{2h},
\end{equation}
where we have introduced $\xi = 1 +\mu Q^* +\eta^*$ and $y=1-\gamma \sigma^2 \chi$. The latter can be shown to be always non-negative $y \ge 0$ as confirmed by numerical inspection. Since we want $x^*>=0$ we need $\xi>0$ to have at least one feasible solution. Given that, and the fact that $\sqrt{(h\xi-y)^2+4h\xi}>|h\xi-y|$ we find that the only non-trivial stationary solution is the one with the plus in Eq. (\ref{eq:second_order_sol}). We also want to consider the case in which $h\rightarrow 0$ where Eq. (\ref{eq:second_order_sol}) is ill-defined when we take the plus sign. To regularize it, it is sufficient to multiply numerator and denominator by $h\xi-y-\sqrt{(h\xi-y)^2+4h\xi}$. The final result for the stationary solution is
\begin{equation}
 x^* = \frac{-2\xi}{h\xi-y-\sqrt{(h\xi-y)^2+4h\xi}} H[\xi].
 \end{equation}

 \section{Supplementary Figures}\label{app:Supp_Figures}
\begin{figure}
    \centering
    \includegraphics[width=1\linewidth]{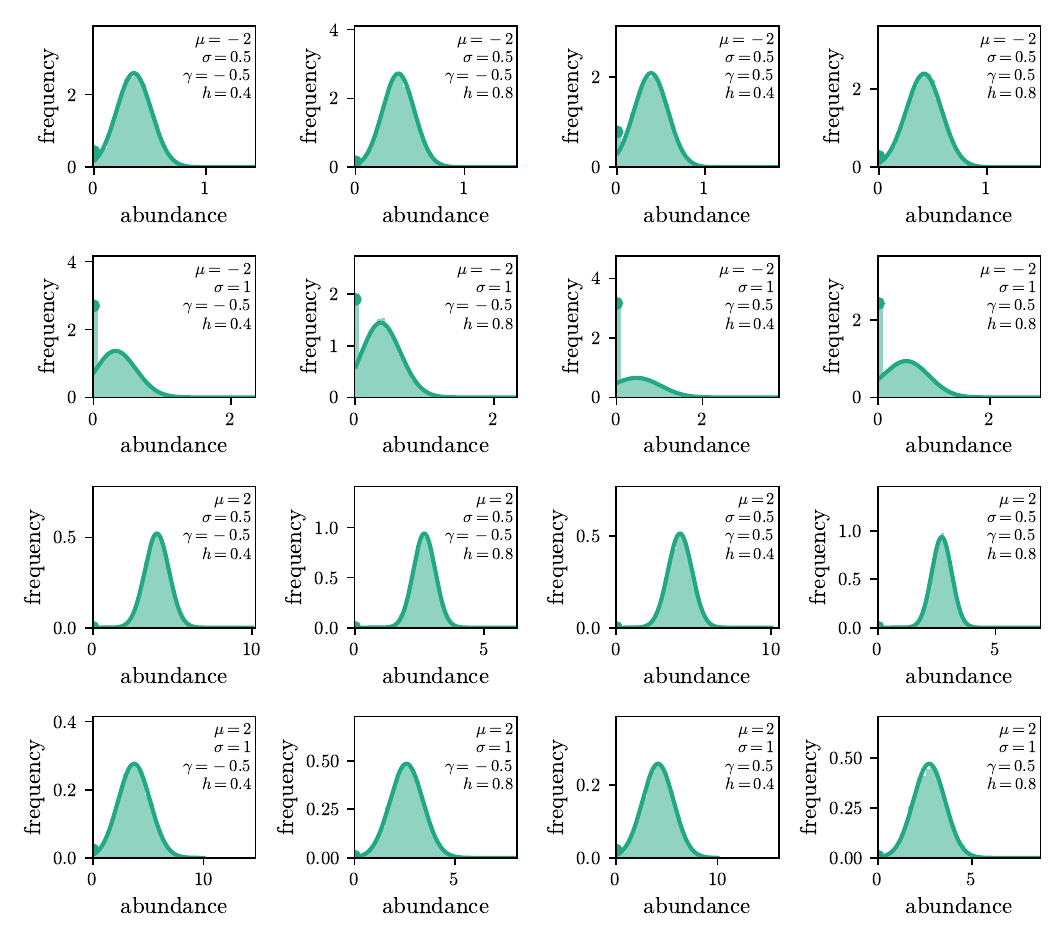}
    \caption{Species Abundance Distribution for different combinations of the control parameters of the model. Comparison between the theoretical species abundance distribution and the histogram of the stationary samples obtained from $10$ simulations with $1000$ species. The values $x^*$ from the simulations are averaged over the last $5\%$ of the trajectories that last for a total simulation time of $100$.}
    \label{fig:SI_SAD}
\end{figure}

\begin{figure}
    \centering
    \includegraphics{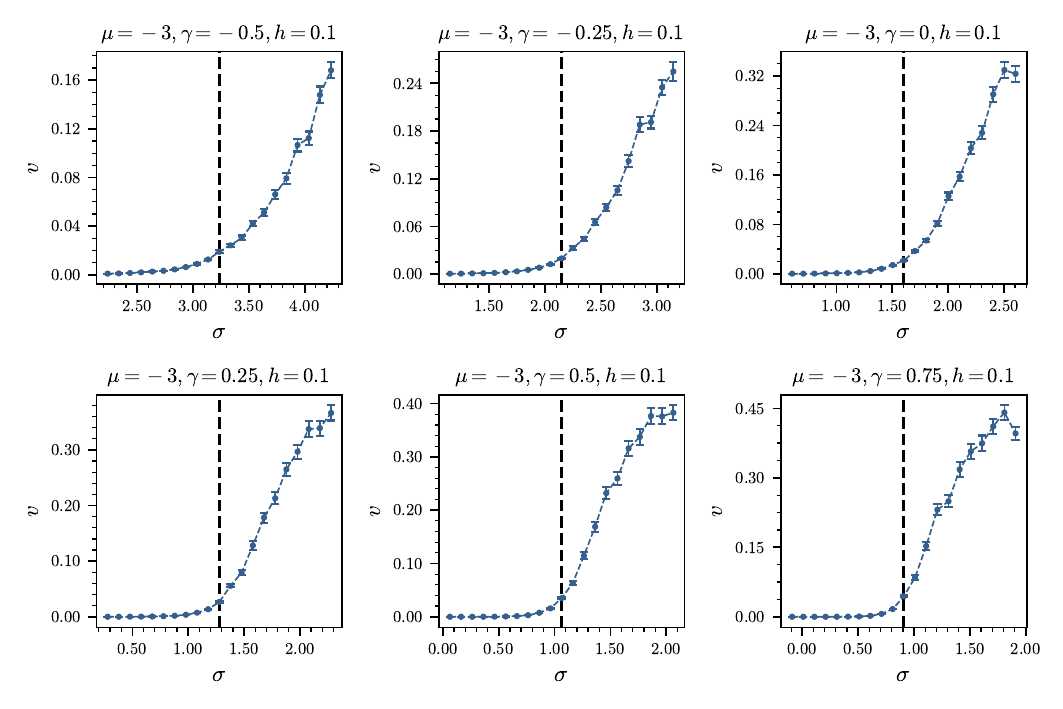}
    \caption{Numerical validation of critical-point prediction. Comparison between the critical interaction strength $\sigma$ predicted by solving Eq.~(\ref{eq:stability_condition}) self-consistently for different values of the symmetry parameter $\gamma$, with $\mu=-3$ and $h=0.1$ held fixed. Numerical validation is performed by computing the volatility parameter $v$ defined in Eq.~(\ref{eq:v_def}). Each point is obtained as the average over $225$ realizations of the interaction matrix with $1000$ species.}
    \label{fig:supp_critical_condition}
\end{figure}

\begin{figure}
    \centering
    \includegraphics{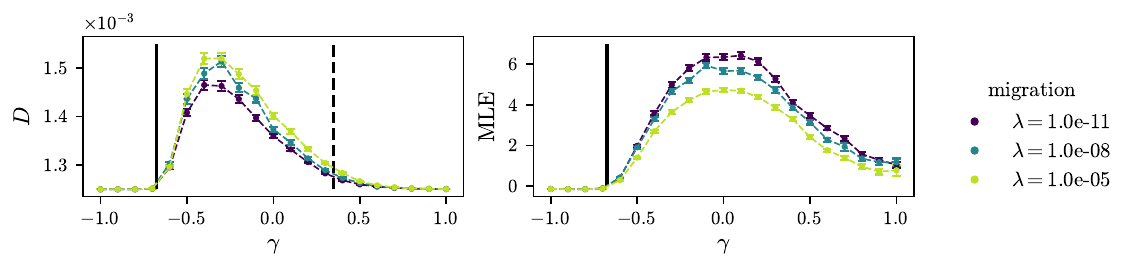}
    \caption{Comparison of the Dimension of activity (D) and Maximum Lyapunov Exponent (MLE) evaluated along a projection of the phase diagram, obtained fixing $\mu = -3$, $\sigma = 5$, $h = 0.1$ and varying $\gamma$, for different values of the migration term $\lambda$. Each point is obtained as the average over $225$ realizations of the interaction matrix with $800$ species. The solid line marks the separation between the Unique Fixed Point phase and the Multiple Attractors phase and is determined by the self-consistent condition Eq.(\ref{eq:stability_condition}). The dashed line marks the separation between the qualitatively different behaviors in the multiple attractors phase and is determined approximately.}
    \label{fig:supp_migration}
\end{figure}

\begin{figure}
    \centering
    \includegraphics[width=0.5\linewidth]{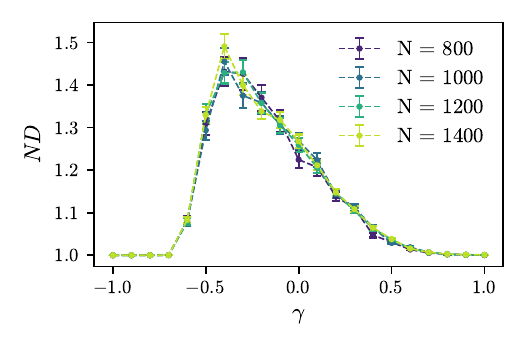}
    \caption{Dependence of the non-normalized Dimension of Activity $ND$ on the number of species $N$ as a function of the interaction symmetry parameter $\gamma$.}
    \label{fig:supp_N}
\end{figure}

 In this appendix we present a more detailed comparison between theoretical predictions and numerical simulations for different parameter choices of the model, focusing on the SAD in Fig.~\ref{fig:SI_SAD} and on the transition between the Unique Fixed Point and Multiple Attractors in Fig.~\ref{fig:supp_critical_condition}. We also investigate the effects of the migration term $\lambda$ and the total number of species $N$ on the computation of the dimensionality of activity, respectively in Fig. \ref{fig:supp_migration} and Fig. \ref{fig:supp_N}.

\end{document}